%% file: top.tex
\RequirePackage{fix-cm}
\documentclass{article}          % twocolumn
\usepackage{graphicx}
\usepackage{mathptmx}      % use Times fonts if available on your TeX system

\usepackage{verbatim}    % Added by AES from GlobalSIP
\usepackage{amsmath}     % Added by AES from GlobalSIP
\usepackage{amssymb}     % Added by AES from GlobalSIP
\usepackage{bm}     % Added by AES for bold in Math
\usepackage{picins}     % Added by AES for bio pics

\usepackage{fancyhdr}

\newcommand{\mxstm}{\ensuremath{\mathit{STM}}}   % Added by AES from GlobalSIP
\newcommand{\mxcard}[1]{\left|{#1}\right|}

\begin{document}

\title{Reconfigurable Digital Channelizer Design Using Factored Markov Decision Processes}

\author{
  Adrian Sapio$^1$\\
  \texttt{asapio@umd.edu}
  \and
  Lin Li$^1$\\
  \texttt{lli12311@umd.edu}
  \and
  Jiahao Wu$^1$\\
  \texttt{jiahao@umd.edu}
  \and
  Marilyn Wolf$^2$\\
  \texttt{wolf@ece.gatech.edu}
  \and
  Shuvra S. Bhattacharyya$^{1,3}$\\
  \texttt{ssb@umd.edu}
}

\date{%
    $^1$University of Maryland, College Park, Maryland, USA\\%
    $^2$Georgia Institute of Technology, Georgia, USA\\%
    $^3$Tampere University of Technology, Finland\\[2ex]%
    %\today
}

\maketitle

% correct bad hyphenation here
\hyphenation{channel-izer chan-nelizer recon-figurable recon-figuration factor-ization}

\begin{abstract}
\begin{sloppypar}
In this work, a novel digital channelizer design is developed through the use
of a compact, system-level modeling approach. The model efficiently captures
key properties of a digital channelizer system and its time-varying operation.
The model applies powerful Markov Decision Process (MDP) techniques in new
ways for design optimization of reconfigurable channelization processing.  
The result is a promising methodology for design and implementation of digital
channelizers that adapt dynamically to changing use cases and stochastic
environments while optimizing simultaneously for multiple conflicting
performance goals. The method is used to employ an MDP to generate a runtime 
reconfiguration policy for a time-varying environment. Through extensive 
simulations, the robustness of the adaptation is demonstrated in comparison 
with the prior state of the art.
\end{sloppypar}

% \keywords{Channelizer \and Markov Decision Processes \and 
% Reconfigurable Architectures \and Software Defined Radio}
\end{abstract}

\thispagestyle{fancy}
\fancyfoot{}
\lfoot{\textbf{\textit{This article has been accepted for publication in a future issue of the Journal of Signal Processing Systems, but has not been fully edited. Content may change prior to final publication.}}}

\section{Introduction} 
\label{sec:intro}
\input{s01-intro}

\section{Background and Related Work}
\label{sec:related}
\input{s02-related}

\section{Reconfigurable Channelizer Design}
\label{sec:application}
\input{s03-application}

\section{MDP-Based Channelizer Control}
\label{sec:method}
\input{s04-method}

\section{Results}
\label{sec:results}
\input{s05-results}

\section{Conclusions and Future Work}
\label{sec:conclusion}
\input{s06-conclusion}

\section{Acknowledgements}
\input{s07-ack}

\bibliographystyle{spmpsci}      % mathematics and physical sciences
%\bibliography{refs,refs/refs-missing}   % name your BibTeX data base
\bibliography{refs}   % name your BibTeX data base

%\section{Authors}
%\input{s07-biographies}

\end{document}

%% file: s01-intro.tex
\begin{sloppypar}
Digital channelizers are critical subsystems in wireless communication
systems that are employed when a multiplexed signal contains information
in different frequency subbands, and the application requires separating one
input signal (containing multiple subbands) into one or more output signals
(each containing a subset of the input subbands)~\cite{vaid1993x1}. This 
function is commonly required in cognitive radio systems~\cite{dara2010x1}. 
\end{sloppypar}

\begin{sloppypar}
In this work, we seek to leverage the reconfiguration capabilities of modern
embedded platforms to develop digital channelizers that can better adapt to the
environment in which they are operating. Adapting to the environment using an
effective {\em system-level reconfiguration framework} ({\em SLRF}) can help
these systems operate more effectively --- e.g., with improved trade-offs among
achievable data rate, latency, and energy efficiency.  For this purpose, we
apply Markov Decision Processes (MDPs) in novel ways to make dynamic decisions
on maintaining or adapting signal processing configurations during channelizer
operation.  We propose an MDP-based SLRF to develop dynamic reconfiguration
policies for use in stochastic environments in which adaptation of
hardware/software configurations for digital channelizer processing is
strategic. 
\end{sloppypar}

While the SLRF techniques are developed in this paper with a
specialized focus on digital channelizer implementation, we believe that the
underlying MDP techniques are applicable across many other types of embedded
signal processing systems (ESIPs). Exploring the generalization of our SLRF for
broader classes of ESIPs is therefore a useful direction for future work.

Our MDP-based approach for digital channelizer design optimization results in
increased robustness when used to periodically re-optimize the system policy
specifically for the external environment it is being used in. This periodic
re-optimization can be done completely autonomously by an embedded signal
processor, without any need for human-in-the-loop intervention. The information
our design optimization methods require is completely observable by the system
at runtime. 

The remainder of the paper is organized as follows. We provide a cursory review of the history of channelizers and MDPs, and their development in Section~\ref{sec:related}.
In Section~\ref{sec:application}, we detail the signal processing application and the algorithms involved. In Section~\ref{sec:method}, we introduce our MDP-based approach and illustrate how it is applied to the signal processing application. We follow that in Section~\ref{sec:results} with a summary of the simulations performed and the resulting data and observations that were made. We conclude in Section~\ref{sec:conclusion} with a discussion of future work on the use of MDPs in channelizers.

%% file: s02-related.tex
\begin{figure} 
\centering 
\includegraphics[width=3in]{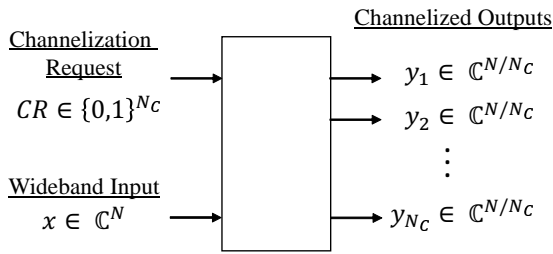} 
\caption{Channelizer inputs and outputs.} 
\label{fig:channelizer_io} 
\end{figure}

A digital channelizer can be generalized as having the inputs and outputs shown
in Figure~\ref{fig:channelizer_io}. Without loss of generality, we represent the
inputs and outputs as frame-based vector quantities, with time decomposed into
fixed-width slots referred to as {\em frames}. The frame arrival rate is
constant and the stream of incoming frames is never ending.  A channelizer is
often a subsystem of a larger signal processing system. For each frame of data,
the channelizer is commanded by higher-level elements of the larger signal
processing system on a per-frame basis. These higher
level elements determine which sub-channels need to be
produced and which do not. 

An example of such a channelizer framework can be found in the cognitive radio
of~\cite{lee2014x1}. In that application, a channelizer is used to isolate
sub-bands within some wireless spectrum dynamically. This dynamic behavior
involves consuming a wideband signal, and applying digital filters and
rate-changing operations to produce an output that contains some subset of the
input signal frequencies. 

In Figure~\ref{fig:channelizer_io}, for each frame $n$ of data, $x^{(n)}$ is a length $N$ complex vector of the wideband input signal. This data is presented to the channelizer alongside
$\mathit{CR}^{(n)}$, a length $N_C$ binary vector that provides the
channelization request for that frame. The channelizer outputs $N_C$ parallel output
data vectors,

\begin{equation}
{y_{\alpha}}^{(n)}, \alpha = \{1,2,\ldots,N_C\}. 
\end{equation}

Each of these vectors contains a channelized subset of the input.

\begin{sloppypar}
Good surveys of popular digital channelizer architectures to date are 
found in~\cite{hu2015x2,vaid1993x1,zhou2006x3}. The most common 
architectures are based on the Cosine Modulated Filter Bank (CMFB), 
Discrete Fourier Transform Filter Bank (DFTFB) and Per-Channel Filter 
Bank (PCFB). Aside from these well-established architectures, several 
other interesting designs for application-specific channelizers can be 
found in~\cite{harr2011x1,harr2013x2,dhab2013x1,edis2012x1}. 
\end{sloppypar}

As illustrated in~\cite{dara2010x1,abu2004x1}, the channelizer is often one of 
the most computationally intensive and power consuming blocks of 
cognitive radio transceivers, mainly due to its need to run at the 
highest data rates. For this reason, several researchers have sought to 
create channelizer designs where the key parameters that control the 
processing (e.g., filter coefficients, data rates, and subchannel masks) 
are configurable at run-time~\cite{chan2006x2,dara2014x1,devi2013x1}. 
We refer to this class of DSP systems as ``reconfigurable 
channelizers'', and point to this active body of DSP research as 
evidence for the importance of optimizing channelizer processing for 
exactly what is required, and nothing more. The goal is generally to 
improve efficiency by increasing processing productivity, while 
simultaneously decreasing energy consumption. 

\begin{sloppypar}
The body of prior work referenced above provides a number of efficient 
channelizer designs that can be flexibly configured for different 
trade-offs. However, this body of work does not address how or when the 
configurable parameters are changed, nor provide policies for changing 
them at run-time. In this paper, we develop MDP-based methods to bridge this gap. 
\end{sloppypar}

\begin{sloppypar}
Other researchers have sought to use MDPs with similar goals. 
Wei et al.\ have demonstrated the effective use of 
an MDP to control the processing rate of a network 
router~\cite{wei2015x2}. This work created a Markov model of only the 
external environment, not the system under control. In contrast, as 
described above, our proposed SLRF incorporates Markov models of both 
the controlled system and the external environment, which provides a 
more comprehensive foundation for dynamic adaptation. 
\end{sloppypar}

\begin{sloppypar}
Hsieh et al.~\cite{hsie2014x1} devise a scheduling policy that selects 
among alternative implementations of common functions, such as FFTs. The 
alternative options accomplish functionally the same operation, but with 
different execution times, power demands, and hardware requirements. As 
in our SLRF, Hsieh's approach uses an algorithm to make reconfiguration 
decisions based on what requests are placed on the system at runtime. 
However, in Hsieh's approach, these requests are converted to a time 
series signal, smoothed using a moving average filter, and then compared 
to thresholds in order to derive reconfiguration decisions. The designer 
must commit to a smoothing factor on the incoming requests, and 
effectively assume a-priori some of the resulting dynamics of the 
system. 
\end{sloppypar}

\begin{sloppypar}
Compared to Hsieh's methods, our SLRF takes a very different approach by 
transforming both the system and operating environment into stochastic 
models, which can then be reasoned upon within the framework of MDPs. In 
contrast to the approach of Hsieh, there are no a-priori trade-offs on 
the smoothing of incoming requests. Furthermore, instead of condensing 
the observable data into one-dimensional signals, larger conditional 
probability tables are maintained. Thus, the algorithms in our SLRF can 
incorporate more knowledge into the decision framework. By incorporating 
historical transition probabilities, the MDP is able to infer in 
real-time whether a new request is likely to be the start of an event 
that should be acted upon, or is more likely a spurious request that is 
better ignored. This inference can be performed immediately and without 
the delay associated with the step response through a smoothing filter. 
\end{sloppypar}

As described in Section~\ref{sec:intro}, we apply MDPs as a core part of 
our proposed methodology for reconfigurable channelizer design. A preliminary 
version of this work 
was published in~\cite{sapi2016x1}. This preliminary version built on the
results of~\cite{beni1999x1}, where MDPs were demonstrated to be useful 
tools for controlling resources in  computing systems. 
In our preliminary version~\cite{sapi2016x1}, we introduced two innovations that 
significantly enhanced the 
effectiveness of MDPs for channelizer design optimization. First, we added 
a mechanism to address hardware/software codesign scenarios that involve 
multidimensional design objectives and constraints, which are commonly 
encountered in transceiver system design. This was done through a 
multidimensional framework for the definition of the MDP rewards 
function. 

\begin{sloppypar}
Second, we introduced {\em transition states} in our MDP formulation to 
represent intermediate states (between distinct channelizer 
configurations) in the target system. We applied transition states in 
scenarios where commanding a state change can result in one or more time 
steps (frames) where the system is in a non-productive transition mode. 
Since being in transition from one state to another can result in 
missing real-time deadlines for processing requests, the control policy 
must choose carefully when to command a transition, and only seek to do 
so when the end result will be a net positive for the system in the long 
run, in spite of any short-term negative effects due to the transition 
frames. Such incorporation of transition states within our SLRF extends 
its utility to a broader class of applications, including channelizers, 
where transitions between productive states must be taken into account 
for accurate assessment and optimization of dynamic reconfiguration 
control. To the best of our knowledge, this was the first time 
that transition states and MDPs have been used together in this way in 
reconfigurable embedded systems. 
\end{sloppypar}

\begin{sloppypar}
In this paper, we build on the preliminary version~\cite{sapi2016x1} in three
ways. First, we apply a methodology developed in~\cite{bout1995x1} to transform
an MDP into a {\em factored} MDP. This concept addresses a problem that
frequently occurs with MDPs --- the number of possible states of the model can
be extremely large. As detailed in~\cite{siga2010x1}, a major motivation behind
factored representations is that some parts of this large state space 
generally do not
depend on each other and that this independence can be exploited to derive a
more compact representation of the global state. In our work, 
factorization serves to reduce the storage size of the MDP model and execution
time of the policy generation algorithms. Such advancements are critical
enablers for a future direction of this work --- deploying the modeling
framework and policy generation algorithms to the targeted embedded system.
When the framework and algorithms are integrated with the application on the
embedded platform, they can be used to perform periodic re-optimization of the
reconfiguration policies in addition to applying the policies to manage system
configurations. To be practical in resource constrained and power constrained
embedded environments, the deployed implementations of the modeling framework
and policy generation algorithms must be carefully optimized so that they
consume minimal amounts of storage and impose minimal computational burden. Our
application of factored MDP techniques in this paper is an important step
towards these objectives.

\end{sloppypar}

Second, we detail the findings of an expanded performance analysis of the
proposed methodology. Specifically, we describe a suite of competing control
policies and compare them objectively with the MDP based techniques. The
results show that the MDP based techniques outperform the alternative schemes
in nearly all cases.

Third, we perform a trade-off analysis of the costs and benefits of including
transition states in the framework. This exploration details and quantifies the
design time modeling costs of transition states in both storage size and
execution time. These costs are then contrasted with the benefits in the form
of the run-time performance when transition states are included versus when
they are not. While transition states were introduced in the preliminary
version~\cite{sapi2016x1} as a novel technique for MDP-based design of
reconfigurable embedded systems, no experimental investigation of their
associated trade-offs was provided due to space limitations.  In this paper, we
provide a more complete presentation of transition states by developing such an
experimental study.

%% file: s03-application.tex
In this section, we present a reconfigurable digital channelizer design that
forms the foundation for our MDP-based, adaptive channelization system, which
we present in Section~\ref{sec:method}, and demonstrate experimentally in
Section~\ref{sec:results}.

\begin{sloppypar}
Our channelizer system is implemented on 
the Silicon Labs EFM32GG, a small and low power ARM Cortex M3-based 
microcontroller. The processor is running on the EFM32 STK3700 
development kit, which houses the CPU as well as sophisticated energy 
monitoring circuitry. For this hardware, a channelizer width of $N_C = 
8$ sub-channels is used in an illustrative experiment. 
\end{sloppypar}

\begin{sloppypar}
This particular channelizer system is developed with applicability to wireless
sensor networks, which impose challenging constraints on energy
consumption and resource utilization. However, with its foundation
in MDP techniques, our design methodology is not specific to any particular
domain of channelization applications. For example, the 
methodology can be adapted to large scale, high performance
channelization scenarios that involve dozens or hundreds of sub-channels that
require the use of FPGAs or GPUs to run in real-time. 
Developing such adaptations for these additional classes of processing platforms is an interesting area for future investigation.
\end{sloppypar}

To examine the ability of the system to adapt to its environment, we consider
two separate use cases, which we refer to as $A$ and $B$. Additionally, we
create multiple scenarios within those use cases, by varying parameters of the
application that are understood to be time-varying.  We design two separate
channelizers, one ideally suited for each use case, as detailed in
Section~\ref{sec:DFTFB} and Section~\ref{sec:DCM}. We then employ a
reconfiguration policy derived using our SLRF with the decision-making
authority to select which channelizer algorithm to use at any given time.
Additionally, the algorithms contain configuration parameters, and we give the
SLRF control of these parameters. This results in a unified controller for
reconfiguration, dynamic power management, and online parameter optimization. 

\subsection{Polyphase DFT Filter Bank}
\label{sec:DFTFB}

Use Case $A$ is the application in~\cite{lee2014x1}. In that system, the 
requests are modeled as i.i.d.\ (independent and identically distributed) 
Bernoulli across both the time and sub-channel dimensions. These 
statistics for the requests mean that there is no opportunity to 
anticipate the request vector. For such an environment, a sensible 
option is a filter bank that outputs all subchannels at all times, in 
the most efficient way possible. For this, we use a Polyphase implementation
of the canonical Discrete Fourier Transform Filter Bank (DFTFB)
described in~\cite{vaid1993x1}. 

\begin{figure}
\includegraphics[width=3in]{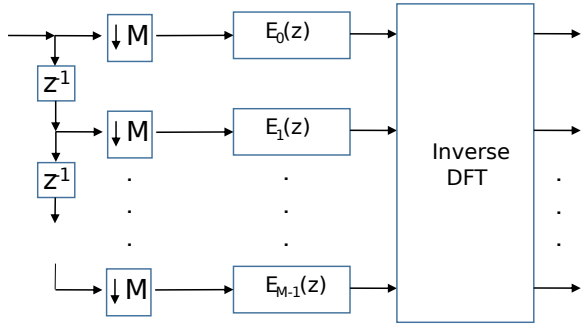} 
\caption{DFTFB block diagram, ${M = N_C}$.} 
\label{fig:dftfb_block_diagram} 
\end{figure}

\begin{figure} 
\centering
\includegraphics[width=3.5in]{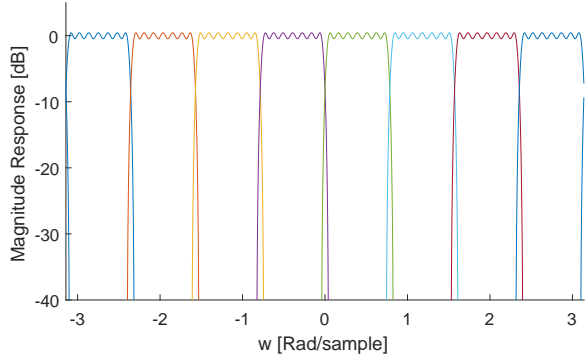} 
\caption{DFTFB magnitude responses.} 
\label{fig:dftfb_mag_responses} 
\end{figure}

To implement this DSP block, we begin by designing a low pass filter to be used
as the ``prototype'' filter in the filter bank. The filter has a passband width
of one eighth of the full spectrum, since there are eight equally spaced
channels. The filter coefficients are chosen using the Equiripple FIR design
method detailed in~\cite{oppe1999x1}.  The prototype filter is then shifted in
frequency, decomposed into its polyphase components $E_m(z)$, and implemented
into the DFTFB, as described in~\cite{vaid1993x1}. A block diagram of
the derived DFTFB is shown in Figure~\ref{fig:dftfb_block_diagram}. The
resulting magnitude response for each of the 8 outputs is shown in
Figure~\ref{fig:dftfb_mag_responses}.

As can be seen from the magnitude responses of the 8 channelized 
outputs, this filter bank can simultaneously channelize all of the 
sub-channels, and thus, we require no tunable parameters for this 
algorithm. In order to optimize for bursts of communication activity as 
well as idle time, we give the controller the ability to put the DFTFB 
in and out of a {\em sleep mode}. The DFTFB remains resident in the current 
configuration, and can be gated on and off very quickly. 
The gating off of the DFTFB corresponds to its sleep mode.

\subsection{Tunable Polyphase Decimation Filter}
\label{sec:DCM}

Use Case $B$ is the Sequential Sensing application in~\cite{xu2011x2}, where a
channelizer with the same inputs and outputs as Use Case $A$ is required.
However, the request statistics imposed on this channelizer are quite different
from those in Use Case $A$. In Use Case $B$, the channelizer is requested to
produce only one output subchannel at a time.  One or more frames (usually
multiple frames) elapse between requests for different subchannels. 

\begin{figure} 
\begin{center}
\includegraphics[width=2in]{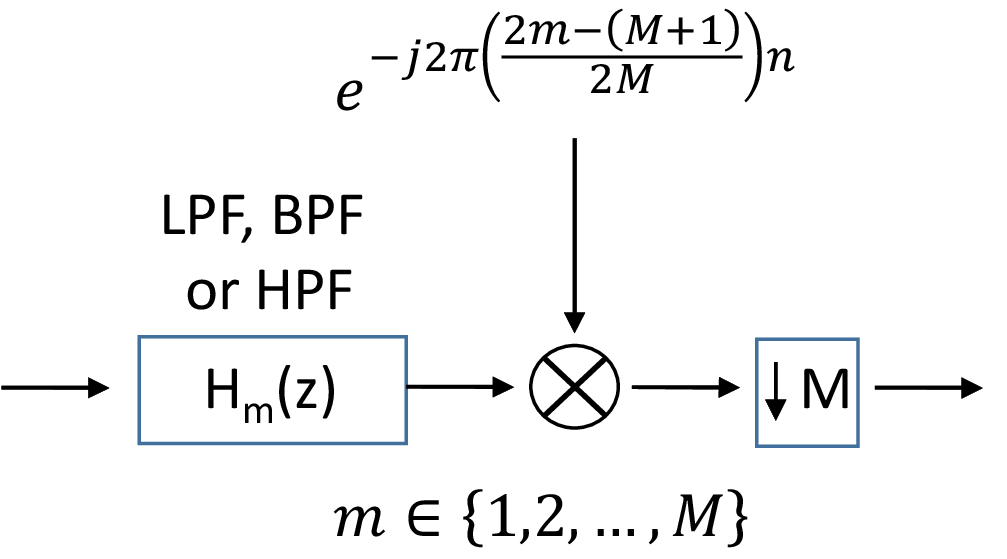} 
\caption{DCM block diagram, ${M = N_C}$.} 
\label{fig:dcm_block_diagram} 
\end{center}
\end{figure}

Since only one channel is requested at any given time, we only need a tunable
decimation of the input data --- i.e., to filter out the unwanted subchannels.
For this, we employ a polyphase implementation of an 8-to-1 decimation (DCM)
filter and mixer as described in~\cite{farz2014x1}, shown in 
Figure~\ref{fig:dcm_block_diagram}.

The operation shown suppresses all but one subchannel out of the incoming
signal, and then uses a complex mixer to shift the extracted channel down to be
centered at DC. Once centered at DC, a simple decimation of samples gives the
resulting output stream. The same filter coefficients used for the prototype
low pass filer of the DFTFB can be used in the DCM.  Such a DCM design produces
the same frequency response per subchannel. Prior to implementation, we utilize
the polyphase technique detailed in~\cite{farz2014x1} to reduce the runtime
processing requirements further without changing the resulting filtering
operation. We refer to the resulting subsystem
as a {\em polyphase decimation filter}. 

Unlike the DFTFB, this configuration does have tunable parameters: the filter
coefficients and mixing frequency. Using 8 parameter sets, this algorithm can
be modified to select any of the 8 subchannels, effectively being an efficient
low-pass, band-pass or high-pass decimation filter.  Both the filter
coefficients and the amount of frequency shifting are tunable, as shown in the
block diagram (Figure~\ref{fig:dcm_block_diagram}). The signal is first passed
through a digital filter $H_m(z)$, whose coefficients are specific to each
channel $m$. Then, the filter output is shifted in frequency by multiplying it
with a sinusoidal signal, whose frequency is also specific to each channel $m$.
The formula to generate the sinusoidal frequency is the exponential shown in
the block diagram.  This configuration is also designed to be kept in a sleep
mode during periods of idle user activity. 

%begin revision Adrian 20171115

%Unlike the DFTFB, this
%configuration does have tunable parameters: the filter coefficients and mixing
%frequency. Using 8 parameter sets, this algorithm can be modified to select any
%of the 8 subchannels, effectively being an efficient low-pass, band-pass or
%high-pass decimation filter. This configuration is also designed to be put in
%and out of a sleep mode during periods of idle user activity. 

%end revision Adrian 20171115

\subsection{Summary of Processing States and Their Properties}
\label{sec:states:summary}

Our MDP framework requires an enumeration of the states that the processing
system can be in at any time. Our experimental embedded system has 13 states,
which fall in the categories listed in Table~\ref{tab:sp_table}. 

%begin revision Adrian 20171125

The first row of the table covers the states when the system is in a sleep
mode, with either the DCM or DFTFB ready to run. We make the distinction
between these as two separate states to allow the model to capture any
difference in time that it may take to re-enable the resident and already
initialized algorithm out of sleep mode compared to switching to the other
algorithm. Further discussion on these delays will be presented in
Section~\ref{sec:s04_transition_states}.

%end revision Adrian 20171125

\begin{sloppypar}
The last two states, whose labels are prefixed with ''Trans.'', are states of
being \textit{in transition} to the DFTFB or DCM, respectively. The time
required by the processing system to transition between states is an important
detail in this framework. The incorporation of transition states into the MDP
is a novel contribution in our work that is intended to take such transition
times into account (detailed in Section~\ref{sec:s04_transition_states}).  This
concept of transition states allows an SLRF to compute decision paths involving
transitions that can take multiple time frames to complete. 
\end{sloppypar}

The third column of the table shows the number of channels provided by the system while in each state. Note that while in transition, the system is consuming power but not producing any channelized data. 

The fourth column of the table shows the CPU power consumed by the system in
each state. These measurements were performed at design time by putting the
processor into test modes created for this purpose. Each test mode loaded a
single configuration and iterated at the experimental application's frame rate.
With the processor operating in such a test mode, the Silicon Labs EFM32GG
development tools allowed the power consumption of the associated state at the
associated frame rate to be measured.

\begin{table}[h!]
\centering
\caption{Categories of processing states and their properties.}
\label{tab:sp_table}
\begin{tabular}{cccc}
\hline\noalign{\smallskip}
State & Num & Num Channels & Average \\    
Category & States & Provided     & Power \\
\noalign{\smallskip}\hline\noalign{\smallskip}
SLEEP & 2 & 0 & 5.36 $\mu$W\\
DCM & 8 & 1 & 7.61 mW\\
DFTFB & 1 & 8 & 17.92 mW \\
Trans.\ DFTFB & 1 & 0 & 10.25 mW \\
Trans.\ DCM & 1 & 0 & 10.25 mW \\
\noalign{\smallskip}\hline
\end{tabular}
\end{table}

It is clear from Table~\ref{tab:sp_table} that the DFTFB is the most productive
configuration (producing all 8 subchannels), while being the most power hungry
in its ON state.  It is also clear from the table that the DCM algorithm
represents a less productive configuration (producing only 1 subchannel)
compared to the DFTFB, but with the benefit of reduced power consumption. If
only one channel is requested for an extended period of time, then a rational
controller should select the DCM configuration over the DFTFB during that time
in order to conserve power. This means the controller must balance the short
term penalty of a non-productive transition with the long term benefit of the
presumably more favorable new state. 

It can be seen from Table~\ref{tab:sp_table} that the number of channels affects
the number of states, and thus, the size of the MDP state space. This 
has significant implications on the resources required to host an MDP-based 
control policy on the target system, and ultimately, on the scalability of this
approach to channelizers with more than 8 channels. This concept will be explored in detail in Section~\ref{sec:factorization}.

%% file: s04-method.tex
\begin{sloppypar}
In this section, we develop an SLRF for modeling reconfigurable channelizers
with the goal of generating run-time control policies that can be steered in
terms of multidimensional operational objectives, including latency,
throughput, and energy efficiency. The procedure is to first create a Markovian
model of the system, and then use an MDP solver to generate a control policy
from the developed system model. We emphasize here that the system and the
environments that it operates in need not be Markovian or even stochastic in
nature, and the Markovian assumptions are made as approximations expressly for
the purpose of arriving at the control policy. These assumptions are validated
by evaluating the resulting control policy on the real system (not the model)
in its intended use case. 
\end{sloppypar}

%begin revision Adrian 20171125

\begin{sloppypar}

\begin{figure} 
\centering 
\includegraphics[width=3in]{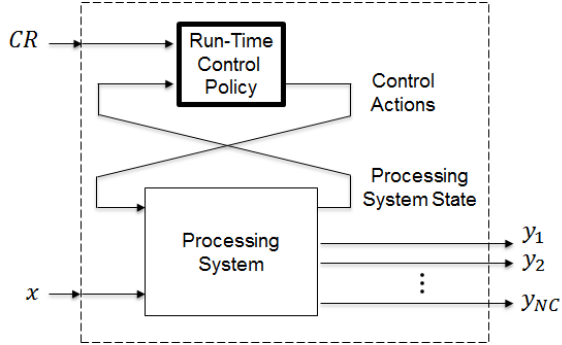} 
\caption{Dynamically reconfigurable channelizer.} 
\label{fig:reconfig_channelizer} 
\end{figure}

The resulting MDP-based dynamically reconfigurable channelizer is illustrated
by the block diagram shown in Figure~\ref{fig:reconfig_channelizer}. The key
feature of this system is that the channelization requests do not have direct
control over the processing system. Rather, the channelization requests go only
to the MDP-generated run-time control policy, which decides when and how to act
on each specific request.  The policy determines the best action to take, 
with the objective of maximizing the long-term average performance rather
than solely based on an immediate reward.
To make this determination, the policy uses models of the application
and processing system characteristics. The policy may decide to
reconfigure the processing system immediately if that is 
assessed as the best decision, or
counterintuitively, it may decide to ignore a request that it predicts is a
spurious request and would not justify a reconfiguration event.

\end{sloppypar}

%end revision Adrian 20171125

The key components of the MDP underlying our reconfigurable channelizer system
are the 4-tuple $(S, A, \mxstm, R)$, where the components of this 4-tuple are
respectively referred to as the system {\em state space}, {\em action space},
{\em state transition matrix} ($\mxstm$), and  {\em reward function}.  The
state space $S$ is defined by enumerating all possible states of the external
requests imposed on the processing system (channelization requests), as well as
a list of modes that the processing system can be in at any time
(reconfiguration states), which were detailed in Section~\ref{sec:application}.
The combination (product) of these two subspaces (external requests and
processing modes) yields the state space of the channelizer system.

For the Action Space ($A$), we give the MDP policy control over the
reconfiguration decision, as well as selected parameter values within particular
configurations. As a result, the action space consists of all the possible
configurations and parameter values that can be commanded.  

The $\mxstm$ is
a stochastic matrix that defines the probability of the next state given the
existing state, conditioned on a given action. This matrix is obtained by
multiplying together the independent statistics of the external channelization requests
with the conditional statistics of the processing system's state transitions. 
The statistics of the channelization requests used to generate the $\mxstm$ are given by the following equations.

\begin{equation}
%\label{eq:sp_stm_case2}
P(CR_{j|i}) = \begin{cases}P_0(CR_{j}), & i=i_0\\P_1(CR_{j}), & i \neq i_0 \end{cases},
\end{equation}

\begin{equation}
%\label{eq:stm_app_A_def}
{P_0(CR_j)}= (p_{start}){P_D(CR_j)} + (1-p_{start})1_{\{j=i_0\}}
\end{equation}

\begin{equation}
%\label{eq:stm_app_A_def}
{P_1(CR_j)}= (p_{stop})1_{\{j=i_0\}} + (1-p_{stop}){P_D(CR_j)}
\end{equation}

\begin{equation}
%\label{eq:stm_app_A_def}
{P_D(CR_j)}=\beta^{\sigma(j)}(1-\beta)^{N_C-\sigma(j)}
\end{equation}

\noindent where $i_0$ is the state where no processing requests are incoming
(representing periods of inactivity), $\sigma(j)$ represents the number of
requested subchannels in the CR state $j$, $\beta$ is a parameter used to
simulate various levels of communication activity, and $p_{\mathit{start}}$,
$p_{\mathit{stop}}$ are used to simulate the system entering and exiting
periods of inactivity. The statistics of the processing system used to generate
the $\mxstm$ are detailed in Section~\ref{sec:s04_transition_states}.

\subsection{Multiobjective Rewards} 
\label{sec:rewards_of_system} 

For the reward function $R$, we contribute a methodology for 
incorporating multidimensional design objectives into an MDP-based
channelizer design framework. 
Given a set $X = \{x_1, x_2, \ldots, x_{N_R}\}$ of $N_R$ evaluation 
functions for key performance metrics, a reward function 
$R : (S \times A) \rightarrow \mathbb{R}$
is defined in terms of these metrics for each action in each 
state. Here, $\mathbb{R}$ denotes the set of real numbers. 

Each evaluation function $x_i : (S \times A) \rightarrow \mathbb{R}$ is used to
estimate system performance in terms of a specific implementation concern, such
as average energy consumption, latency, or throughput. These estimation
functions can be formulated at design time by using knowledge of the system and
its available configurations, or measured online by supporting instrumentation.
The result of each evaluation function $x_i$ is transformed by a mapping $g_i:
\mathbb{R} \rightarrow [0,1]$, which is defined at design time for each metric.
These transformations are introduced to normalize the performance metrics in
order to allow them to be combined into the single scalar output of $R$.  This
kind of transformation and combination follows the \textit{scalarization}
approach to multiobjective optimization, as described in~\cite{bjor2014x1}. 

The combination of the transformed results of the evaluation functions 
are performed by a set of weights  $\rho $ $= \{r_1, r_2, \ldots, 
r_{N_R}\}$, one corresponding to each metric, such that 

\begin{equation} 
\label{eq:scalarization_weights_def} 
(r_i \in [0,1] \mathrm{\ for\ each \ } i) \mathrm{\ and \ } (1 = 
\sum_{i=1}^{N_{R}}{r_i}). 
\end{equation} 

\noindent Determining these weights $\rho$ is a design time aspect of our SLRF.
The weights are determined once and then continually used to steer any
executions of the solver to seek policies that achieve the desired
prioritization of metrics in consideration with the observed external
environment statistics. 

Once the evaluation functions $X$, transformations $\{g_i\}$, and 
combination weights $\rho$ are determined, the reward function can be 
evaluated using Equation~\ref{eq:r_def} for any given $s \in S$ and 
$a \in A$.

\begin{equation} 
\label{eq:r_def} 
R(s,a)=\sum_{i=1}^{N_{R}}{r_i}\,{g_i(x_i(s,a))} 
\end{equation} 

In our experiments, we define the rewards as follows. First, we define 
$g_1$ as the normalized rate of successful channelization requests. This 
can be expressed as $(\eta_r - \eta_d) / N_C$, where $\eta_r$ represents 
the total number of channelization requests input to the system during a 
given time interval $\tau$, and $\eta_d$ represents the number of 
dropped requests (i.e., requests where there is a failure to produce the 
desired channel) during $\tau$. 

We define $g_2$ based on a formulation in~\cite{wei2015x2} for the {\em 
normalized power savings} of an electronic system. Specifically, in order to
normalize power consumption and treat it as a form of savings, we measure power
consumption ($x_2$) in each state and note the minimum and maximum possible values. 
Then we transform the power measurement relative to the maximum and minimum power 
that the system consumes in all of the possible states ($g_2$). The result is shown in Equation~\ref{eq:g2_def} and Equation~\ref{eq:x2_def}.

\begin{equation}
\label{eq:g2_def}
g_2(x_2(s,a)) = \frac{x_{2, \mathit{MAX}} - x_2(s,
a)}{x_{2, \mathit{MAX}}-x_{2, \mathit{MIN}}},
\end{equation}

\noindent where

\begin{equation}
\label{eq:x2_def}
\begin{split}
x_2(s,a)\equiv \mathit{Power\ Consumed}(s,a) \\
x_{2, \mathit{MAX}} =  \max_{s^\prime,a^\prime}\{x_2(s^\prime,a^\prime)\} \\
x_{2, \mathit{MIN}} =  \min_{s^\prime,a^\prime}\{x_2(s^\prime,a^\prime)\}. \\
\end{split}
\end{equation}

Note that this definition is consistent with the convention we have defined:
the most power hungry state has $g_2=0$ (and thus is the least rewarded), while
the least power hungry state has $g_2=1$ (and thus is the most rewarded). 

The combination of rewards functions $g_1$ and $g_2$ effectively steer the MDP
to find policies that are most productive at channelizing the incoming signal
as per the channelization requests, while consuming as little power as possible
on average.

\subsection{MDP Solver and Policy} 
\label{sec:policy} 

With the definitions and rewards described above, an off-the-shelf MDP 
solver can be employed to generate a policy that simultaneously seeks to 
maximize the rate of successful channelization requests while consuming 
the least energy possible, taking into account both the physical 
characteristics of the processing system as well as the independent 
statistics of the operating environment at the current time. 
In our experiments, we apply the open source solver 
MDPSOLVE~\cite{fack2011x1} in MATLAB.
 
The resulting control policy has the form $f: S \rightarrow A$ --- i.e., a
mapping from states into actions.  This mapping can be implemented as a
function or simple lookup table that is invoked or accessed once per frame,
respectively. To execute the controller, the incoming request is combined with
the current processing system state. The result is then used as an index to
lookup the operations involved in the next optimal control action. 

In this example application, the total number of states is 3328 and the total
number of actions is 13. For these quantities, the action can be encoded into 4
bits and thus 2 encoded actions can be packed into 1 byte of storage. The
result is a policy that can be packed into 1.6kB. For our prototype hardware
implementation, it was feasible to simply store the policy as a lookup table in
RAM and index it to look up the next action. 

%begin revision Adrian 20171125
%switched the order of subsections "Factorization" and "Transition States"

\subsection{Transition States}
\label{sec:s04_transition_states}

\begin{sloppypar}
In our design context, the processing system is typically a deterministic,
controllable machine, such as a general purpose processor (GPP), programmable
digital signal processor (PDSP), field programmable gate array (FPGA) or
graphics processing unit (GPU). Our framework assumes that this type of
processing system can be modified or reconfigured through the action decision
of the MDP. By definition, in MDP frameworks the system is assumed to transition
probabilistically from one state to another as a result of an action decision.
This abstract probabilistic transition viewpoint is not immediately amenable to
modeling the transitions of a deterministic processing machine.  Rather, the
resulting state changes in the processing system are better described as a
change that is guaranteed to occur but can take some fixed or variable amount
of time to complete. Additionally, the change may take longer than a single
frame to complete. Some examples of the types of operations typically
encountered in this context that must be accounted for are: (1) computation of
the schedule for a dataflow graph before being able to execute it, (2)
allocation of memory from an operating system heap when initializing algorithms,
(3) the block copy of code or data from a slower, larger long-term storage to a
smaller, faster location (e.g., page fault), (4) the block copy of code from
non-executable regions to executable regions (e.g., overlays), and (5) dynamic
full or partial reconfiguration (DPR) of FPGA regions, to name a few.
\end{sloppypar}

To assign the required state transition probabilities
in this context, suppose that
the processing system receives action $w$ in frame $n$ while in state $\mathit{sp}^{(n)}=u$,
and that this state/action pair is known to deterministically transition the processing system
to a new state $v$ in an amount of time denoted as $T_{u,v|w}$, which need not
be an exact multiple of the frame period $T_{F}$. 

If $T_{u,v|w} < T_{F}$, then the conditional State Transition Matrix for the processing system (SP\_STM) is trivially computed by 

\begin{equation}
\label{eq:sp_stm_case1}
\mathit{SP\_STM}_{i,j|w} = \begin{cases}1, & j = v\\0, & otherwise\end{cases}
\end{equation}

\noindent This represents a guaranteed (i.e., with probability 1) transition of the processing system to state $v$ that completes before the start of the next frame. 

If, on the other hand, this transition takes longer
than $T_{F}$, we define a new processing system state $m$,
which is defined as the state of being \textit{in transition} from
$\mathit{sp}=u$ to $\mathit{sp}=v$. In this case, the conditional SP\_STM matrix
is calculated by

\begin{equation}
\label{eq:sp_stm_case2}
\mathit{SP\_STM}_{i,j|w} = \begin{cases}1, & i=j=v\\1,& i=u , j=m\\1-c, & i=j=m\\c, & i=m,j=v\\0 & otherwise\end{cases},
\end{equation}

\noindent where

\begin{equation}
\label{eq:bernouli_param}
c = \left[\mathit{floor}\left(\frac{T_{u,v|w}}{T_{F}}\right)\right]^{-1}.
\end{equation}

For example, if the processing system transition takes 4.67 frames to complete
and the action is held constant until the completion of the transition, then
the system will begin transitioning immediately following the triggering
action, and will remain in transition for 4.67 frames before arriving at the
destination state. In this case, the conditional transition matrix states that
with probability 1, the processing system will transition from the starting
state to the transition state in the first frame, and then for each subsequent
frame will remain in the transition state with probability 3/4, and will jump
to the destination state with probability 1/4. This is exactly how the
transition would appear to an agent who naively observes the processing state
during just the transition sequence. This agent would observe 3 non-transitions
and 1 transition out of 4 trials. 

We can model observations during the transition as a Bernoulli random variable,
as was done in~\cite{beni1999x1} through the use of Bernoulli trials. Here, we
take the two random outcomes as those of remaining in transition and completing
the transition. Then the Maximum Likelihood Estimator (MLE) of the Bernoulli
parameter can be shown to be exactly as given by
Equation~\ref{eq:bernouli_param}.  For this reason, the Bernoulli probability
mass function is given by the corresponding row of the conditional transition
matrix, as expressed in Equation~\ref{eq:sp_stm_case2}.  With knowledge (or an
estimate) of the transition time from each state/action pair in the model, the
entire set of SP\_STM matrices can be populated in this manner.

\subsection{Factorization} 
\label{sec:factorization} 

In this work, the MDP model and solver components are implemented and invoked
at design time, in order to generate a control policy that is used at run time.
However, an interesting future direction for this work is that of transferring
the MDP model and solver to the target system such that the solver can be
invoked periodically at run time.  The solver can then be applied to
dynamically re-optimize the control policy in response to a changing external
environment.  Working towards this goal, in this section we analyze the target
platform resources necessary for embedded deployment of the MDP model and
solver. The main aspects of resource utilization that we investigate here are
(1) the size of the four MDP constructs $(S, A, \mxstm, R)$ that need to be
held in memory, and (2) the execution time of the MDP solver required to
generate the control policy. 

\begin{sloppypar}
In this context, we find significant advantages to adopting the Factored MDP
approach developed in~\cite{bout1995x1}. In that work, knowledge of the
stochastic inter-dependencies between the state space variables are exploited
to reduce both the memory requirements and solver execution time. 
In the remainder of this section, we summarize relevant background on 
MDP factorization, and present details of our proposed application 
of factorization techniques to reconfigurable channelizer implementation.
\end{sloppypar}

To facilitate the factorization of MDPs, the state $s \in S$ is generally
described as an instantiation of a discrete multivariate random variable
$\mathbf{Z} = (Z_1, Z_2, \ldots, Z_{N_Z})$, where each variable $Z_i$ 
takes on values in $\mathit{DOM}(Z_i)$, and $\mathit{DOM}(V)$
represents the set of
of admissible values of the random variable $V$. Then a state becomes a
set of instantiations of the $N_Z$ random variables,
and can be written as a vector $\mathbf{z} \in
\mathit{DOM}(\mathbf{Z})$. The size of the state space is defined
by the cardinality of this set, which we denote as
$\mxcard{\mathit{DOM}(\textbf{Z})}$. Using this approach, the state space of the
channelizer can be represented as:

\begin{equation} 
\label{eq:ss_vars} 
\begin{split}
s &= (\mathit{CR}_1, \mathit{CR}_2, \ldots, \mathit{CR}_{N_C}, \mathit{CF}_1, 
\mathit{CF}_2). \\
\end{split}
\end{equation} 

\noindent Here, $CR_i$ is the channelization request for channel $i$, $CF_1$ is
the top-level processing configuration, and $CF_2$ is the processing
subconfiguration. The benefit of using this scheme is that it enables the
explicit specification of the stochastic inter-dependencies of the variables
within the state space. With this in mind, factored MDPs make use of Dynamic
Bayesian Network (DBN) diagrams~\cite{russ2009x1} to explicitly define and
illustrate these dependencies. 

\begin{figure}
\centering 
\includegraphics[width=3in]{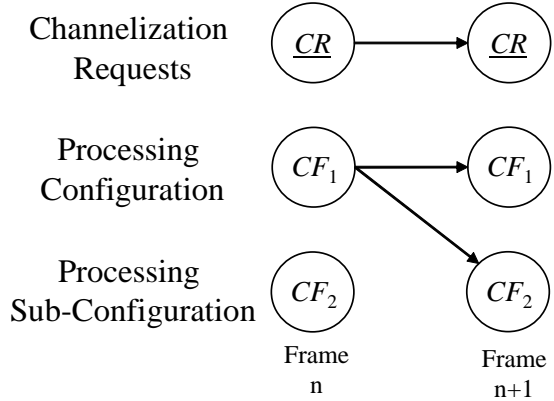} 
\caption{Dynamic Bayesian network representation of the channelizer state space.} 
\label{fig:dbn} 
\end{figure}

A DBN diagram of the channelizer's $\mxstm$ when conditioned on an MDP action
is shown in Figure~\ref{fig:dbn}. Note that the $(\mathit{CR}_1, \mathit{CR}_2,
\ldots, \mathit{CR}_{N_C})$ requests are grouped together into a single vector
$\underline{CR}$ for conciseness. A stochastic dependency between two variables
in the state space (from one time frame to the next) is denoted via the
presence of an arrow between the dependent variables. The absence of an arrow
denotes independence. Thus, the diagram shows that the joint probability
distribution of the channelization requests is dependent only on the requests
in the previous frame, and is independent of the processing configuration. The
processing configuration is dependent only on the previous processing
configuration (since reconfigurations are not instantaneous). However, this
dependency is only on the top-level processing configuration (e.g., DCM, DFTFB,
etc.) and not on the subconfiguration (e.g., the filter coefficients).

\begin{sloppypar}
Knowledge of this underlying stochastic structure within the state space allows
for considerable reduction of the size of the data structures required to store
the MDP model. We highlight the effect on the largest of these components: the
$\mxstm$. Only the conditional probabilities with respect to the dependent
variables need to be stored, rather than with respect to all variables --- as
would be necessary in an equally sized state space where the underlying
stochastic structure is unknown. The factorization made possible by the
knowledge is represented in Equation~\ref{eq:factorization}. The rearrangements
are made possible through (1) independence between the channelization request
and processing configuration, and (2) independence between the channelization
request and the MDP action. 
\end{sloppypar}

\begin{equation} 
\label{eq:factorization} 
\begin{split}
p(s'|s,a) &= p({\underline{\mathit{cr}}}',{\mathit{cf}_{1}}',{\mathit{cf}_{2}}'\:|\:\underline{cr},cf_{1},{cf_2},a) \\
&= p({\underline{\mathit{cr}}}'\:|\:{\underline{\mathit{cr}}})\: p({
\mathit{cf}_{1}}',{\mathit{cf}_{2}}'\:|\:\mathit{cf}_{1},a) \\
\end{split}
\end{equation} 

The resulting reduction in the number of elements in the $\mxstm$ is shown in
Equation~\ref{eq:ss_reduction}. This reduction represents a significant
savings. Note that the quantity shown is the cardinality of the sets, which is
a count of the number of elements regardless of what underlying data type is
used for representation in the MDP model and solver algorithms.  For example,
if the data type used is a 16-bit or 32-bit fixed-point representation, the
total storage size would be 2 bytes or 4 bytes per element, respectively.

\begin{equation} 
\label{eq:ss_reduction} 
\begin{split}
\mxcard{S}^2\:\mxcard{A} &\gg \mxcard{\textrm{DOM}(\underline{CR})}^2+ \\ &\mxcard{\textrm{DOM}(CF_1,CF_2)}\;\mxcard{\textrm{DOM}(CF_1)}\;\mxcard{A} \\
121.8\mathrm{x}10^6 &\gg 66.3\mathrm{x}10^3
\end{split}
\end{equation} 

%end revision Adrian 20171125

%% file: s05-results.tex
\begin{sloppypar}
To evaluate the effectiveness of our {\em MDP-based Reconfigurable Channelizer
System} ({\em MRCS}), we developed a simulation with external requests that
follow the statistics of the two use cases --- here termed $\mathit{IID}$ for
the i.i.d.\ requests of Use Case $A$ (introduced in Section~\ref{sec:DFTFB}), and $\mathit{SEQ}$ for the sequential sensing of Use Case $B$ (introduced in Section~\ref{sec:DCM}). In the following sections we perform three evaluations. 
First, we compare the results against those of manually generated
policies, that we consider representative of a typical approach used in industry. Second,
we compare the results against another method published by researchers. Third, we explore the effectiveness and trade-offs associated with modeling transition states.
\end{sloppypar}

\subsection{Comparison with Manually Generated Policies} 
\label{sec:manual_policies} 

In order to evaluate the effectiveness of the MDP generated control policy, we
created several alternative control policies to compare it against. These are
referred to as the ``manually generated'' policies, and contrasted with the set
of ``MDP generated'' control policies. The manually generated policies were
generated through intuitive heuristics, by first defining common sense rules
for controlling the system in question, and then translating those rules into
code. This represents the traditional method that an embedded software
developer would use to create a reconfiguration policy.  For the manually
generated alternatives, the rules and resulting policies are as follows:

\begin{enumerate}

\item DFTFB --- This policy keeps the DFTFB algorithm on the chip at all times,
and invokes it in all frames regardless of the external requests. This policy
was used purely as a starting baseline, as this policy represents the absence
of reconfiguration options, using the most productive and processor intensive
channelizer available in the system at all times to meet all requests. 

\item DFTFB+Sleep --- This policy also keeps the DFTFB algorithm on the chip at
all times. However, if the number of requested channels is 0, the DFTFB is put
into sleep mode. Otherwise, the DFTFB is kept on. 

\item DCM+Sleep --- This policy keeps the DCM algorithm on the chip at all
times. If the number of requested channels is 0, the DCM is put into sleep
mode. Otherwise, the DCM is kept on and applied to produce one of the
requested channels.  

\begin{sloppypar}

\item DFTFB+DCM+Sleep --- This is a set of policies that use both the DFTFB and DCM algorithms.
The reconfiguration decision occurs based on how many channels are requested in
the upcoming frame. If less than DFT\_THRESH channels are requested, the DCM
algorithm is used. If more than this threshold are requested, the DFTFB
algorithm is used. Additionally, if the number of requested channels is 0, the
algorithm that is currently is loaded is put into sleep mode. If a
reconfiguration is in progress, it is allowed to finish regardless of incoming
requests.  The DFT\_THRESH parameter is varied from 2 to 6, resulting in 5
different control policies. 

\end{sloppypar}

\end{enumerate}

\begin{sloppypar}

In order to compare the policies objectively, we created the following
experimental setup on the EFM32GG development board. Both channelizer algorithms
were implemented in C and stored on the external system FLASH. A MATLAB
simulation was created that produced a time series of channelization requests
having the statistics described in the two use cases $A$ and $B$. The time series output
of the simulation was translated to a C array and stored on the EFM32GG. A test
harness was written on the EFM32GG, which was driven by a periodic timer
interrupt. At the interrupt rate, the next channelization request was pulled
from the stored vector and that channelization request was then
used as an input to our dynamically reconfigurable channelizer system. 
\end{sloppypar}

\begin{sloppypar}
This system was implemented in C and
executed on the EFM32GG. In order to facilitate an objective comparison of
control policies, all of the manually generated policies were stored as Lookup
Tables (LUTs) in addition to the MDP generated policies.  This allowed both the
manually- and MDP-generated policies to be invoked by suitably swapping
out the contents of the LUT. 
\end{sloppypar}

\begin{sloppypar}
As part of the test harness, we incorporated a small amount of diagnostic code
to compute performance objective 1 (productivity) in real-time.  This
computation was performed by comparing the produced channelizer outputs with
the requests. A channelization request that was successfully carried out was
labeled a success. Conversely, a request that was not met was labeled a 
failure (e.g., if the 
processing system was in a reconfiguration state during a frame with
channelization requests in it, or if a configuration was in place that could
not produce enough output channels, etc.). The ratio of the successful outcomes
to the number of requests was used to compute a success rate, which was
used as a measure of system productivity.
The measured productivity results were periodically streamed to a
laptop computer using the ARM on-chip trace functionality, and EFM32GG Single
Wire Output (SWO) port. The streamed output for each case was tabulated and
used for comparison. 
\end{sloppypar}

Metric 2 (CPU power consumption) was measured by using the EFM32GG board's
energy monitoring tools. These development tools allowed a very accurate
current measurement to be taken, showing the exact current drawn by the CPU over
time for each control policy. The total current drawn over the total simulation
time was used to create a single metric for average power consumption. 
Thus, a highly repeatable experimental setup was
applied, where all experimental settings were kept the same from case to case
with the only difference being the control policy being used.  

\begin{figure} 
\centering 
\includegraphics[width=3in]{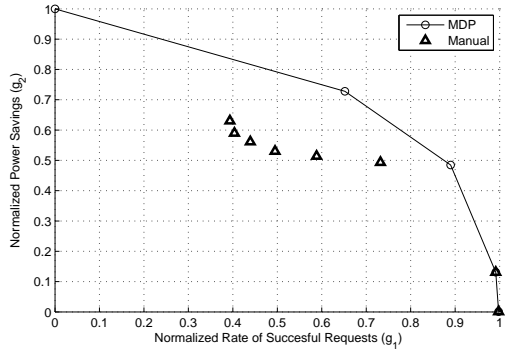} 
\caption{Policy comparison results.} 
\label{fig:results1} 
\end{figure}

Results of our experiments are summarized in Figure~\ref{fig:results1}.
Here, each point in the figure represents the average performance of one policy
over the entire simulation. The MDP policies generated by different
values of $r_1$ are connected together, illustrating a Pareto front generated
by the suite of MDP policies. The manually generated policies are plotted without any connecting lines. If the distance from the origin is used as a scalar metric of
performance, the MDP generated policies all outperform or perform equally to
the best manually generated policies.

\subsection{Comparison with mHARP} 
\label{sec:comparison_mHARP} 

\begin{sloppypar}
Next, we compared our {\em MRCS} to a competing published method, the Highly Adaptive Reconfiguration
Platform (HARP), introduced in~\cite{hsie2014x1}. One modification was needed,
as the published HARP made decisions purely to optimize energy efficiency. This
was inadequate for our setup, as the most energy-efficient result is one where
the system never leaves its sleep state. To remedy this, the single metric in
HARP was replaced with our multidimensional reward framework
(Section~\ref{sec:rewards_of_system}) to construct a useful policy and also to
provide a fair comparison between the two methods. We refer to this modified
method as {\em multiobjective HARP} ({\em mHARP}).
\end{sloppypar}

\begin{figure} 
\centering 
\includegraphics[width=3in]{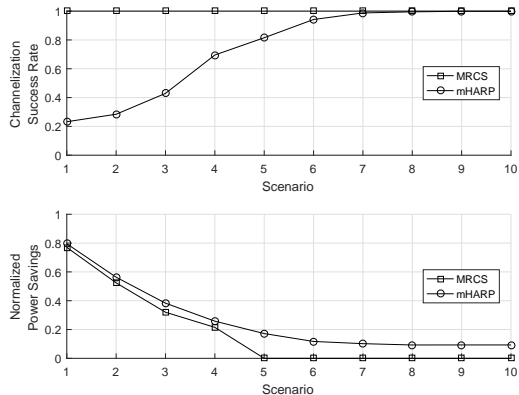} 
\caption{Experimental comparison between MRCS and mHARP, for IID use case.}
\label{fig:combined_results_iid} 
\end{figure}

\begin{figure} 
\centering 
\includegraphics[width=3in]{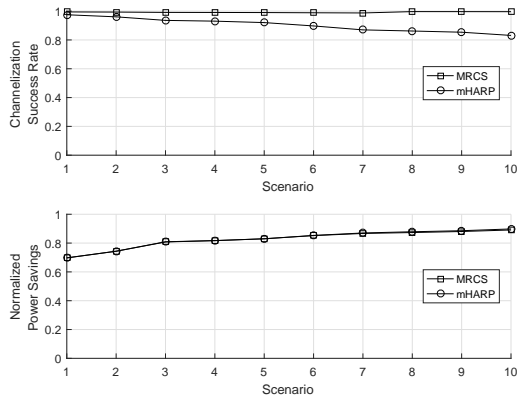} 
\caption{Experimental comparison between MRCS and mHARP, for SEQ use case.}
\label{fig:combined_results_seq} 
\end{figure}

\begin{sloppypar}
For each of the two competing techniques, we created 10 scenarios by varying
the Bernoulli parameter in use case $A$, and another 10 by varying the 
channel dwell time in use case $B$. The result is 20
simulations where our method and the baseline method (described below) were
allowed to implement and run the optimal control policy for the given use case
and external environment. The system characteristics and measurements described
in the previous section were used to define the processing system under
control. The results from our experiments are summarized in
Figures~\ref{fig:combined_results_iid} and~\ref{fig:combined_results_seq}, for use cases
$A$ and $B$, respectively. As previously mentioned, HARP requires
a-priori tuning for a given desired system dynamic. In this simulation, we
optimized mHARP for power savings. The results show that when tuned in this
way, mHARP does well in this metric for all scenarios (producing slightly
better performance than our MRCS approach), but greatly sacrifices performance
in the success rate for half of the scenarios. Conversely, when we attempted to
optimize mHARP for the success rate, we saw large shortcomings in the power
savings. In contrast, MRCS {\em involves no a-priori tuning, and optimizes all decision making for each scenario individually} without compromises. These results show MRCS to have greater robustness to a wide range of parameters in different applications, all without
any human-in-the-loop intervention. 
\end{sloppypar}

\subsection{Trade-offs in Modeling Transition States} 
\label{sec:transition states} 

\begin{sloppypar}
An analysis was performed into the effectiveness of modeling processing state
transitions, as described in Section~\ref{sec:s04_transition_states}.  
Although our  prototype system did not incur large reconfiguration delays, we anticipate larger delays in our future work as we scale up to larger channelizer applications. 
Adding transition states to the MDP model has the undesirable effect of increasing the
size of the state space, which is known to increase the size of the model's
data structures as well as the execution time of the policy generation
algorithms. In order to make informed modeling decisions, it is crucial to
understand what is gained at the expense of these costs.   With these goals in mind, one of the scenarios of the $\mathit{IID}$ application was selected for exploration, and modified in two ways. 
\end{sloppypar}

First, the dynamics of the processing system were modified by changing the
amount of time that transitions of the top-level reconfigurations would take to
complete. This delay was varied between 1 and 5 frames, representative of a
range of a small reconfiguration delay to a large delay. Second, two
alternative MDP modeling approaches were used and compared: one with the
transition states modeled and one without. 

\begin{figure} 
\centering 
\includegraphics[width=3in]{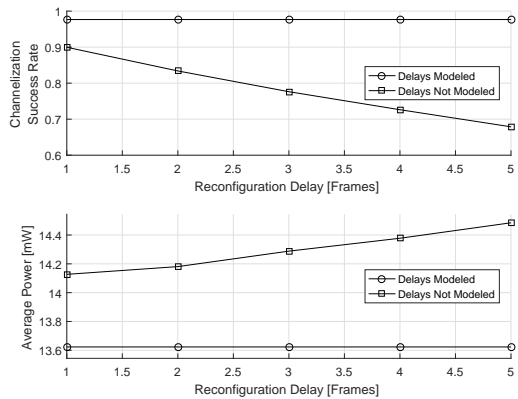} 
\caption{Run-time performance with and without transition delay modeling.} \label{fig:transition_states} 
\end{figure}

The cost of the additional modeling is shown in Table~\ref{tab:ts_comparison}.
The increase in the size of the $\mxstm$ is practically negligible, however the
increase the solver's execution time is not. The benefits of this more
expensive modeling come at run-time, and are shown in
Figure~\ref{fig:transition_states}. This figure shows the resulting 
assessment in terms of the performance metrics defined in the previous section. 

From this assessment, we see that when transitions are not modeled, the
performance of the system (with respect to both metrics) degrades
proportionally with the length of the reconfiguration delays. This degradation
is attributed to the system spending more time in a non-productive
reconfiguration state. In comparison, the MDP that has the transitions modeled
does not exhibit this performance degradation. We attribute these results to
the fact that the MDP with transition states is able to consider the
reconfiguration penalties in its decision criteria, and as a result is more
``reluctant'' to trigger costly reconfigurations.

\begin{table}[h!]
\centering
\caption{Modeling costs with and without transition delay modeling.}
\label{tab:ts_comparison}
\begin{tabular}{ccc}
\hline\noalign{\smallskip}
Delays  & STM Size & Execution Time \\    
Modeled & [Elements] & [Seconds] \\
\noalign{\smallskip}\hline\noalign{\smallskip}
No  & 66020 & 17.2 \\
Yes & 66394 & 24.0 \\
\noalign{\smallskip}\hline
\end{tabular}
\end{table}

%% file: s06-conclusion.tex
\begin{sloppypar}
In this work, we have presented a methodology for design and implementation of
adaptive digital channelizer systems, and we have demonstrated a novel
channelizer design, called the MDP-based reconfigurable channelizer system
(MRCS), that is derived using our new methodology.  Our methodology and MRCS
employ compact, system-level models based on Markov Decision Processes (MDPs)
to generate control policies that optimize the required embedded signal
processing tasks in terms of relevant, multidimensional design optimization
metrics.  Through extensive simulations, we have shown that MRCS outperforms
the prior state of the art in terms of robustness to changing applications and
scenarios.  
\end{sloppypar}

Useful directions for future work include adapting our MDP-based,
reconfigurable channelizer design methodology to derive dynamically
reconfigurable forms of other types or other combinations of channelizer
architectures, and generalizing the proposed design methodology to address
broader classes of embedded signal processing applications.

One requirement of our SLRF is that the statistics of the external environment
and reconfiguration dynamics must be known at design time. In certain
applications, this may not be feasible, or they may be time-varying to such a
point that a policy generated offline at design time may experience a reduction
in effectiveness as these quantities change. 

An important area for future exploration is pairing our framework with learning
strategies to estimate these statistics at runtime for systems where they are
not constant or not known up front. These running estimates could then be used
to periodically re-optimize the control policy and keep it performing optimally
across time-varying use cases and a time-varying environment. 

\begin {comment}

Our SLRF
provides a systematic methodology for dynamic adaptation of embedded signal
processing systems.  The framework includes multiobjective optimization
at its core, embracing the multifaceted nature of signal processing design,
where making strategic trade-offs among conflicting goals is critical. As a
result of this emphasis, the framework can jointly optimize power and
real-time performance, and can readily be adapted to address other
combinations of metrics that are important for a given application. 

The effectiveness of the method was shown in simulation, using empirical 
measurements for the properties of an experimental signal processing system. 
The simulations show the method to outperform the prior state of the art in 
terms of robustness to changing applications and scenarios. 

In this work, we have presented a system level reconfiguration framework 
(SLRF) for using Markov Decision Processes
(MDPs) to generate control policies for embedded systems. Our SLRF
provides a systematic methodology for dynamic adaptation of embedded signal
processing configurations.  The framework includes multiobjective optimization
at its core, embracing the multifaceted nature of embedded systems design,
where making strategic trade-offs among conflicting goals is critical. As a
result of this emphasis, the framework can jointly optimize power and
real-time performance, and can readily be adapted to address other
combinations of metrics that are important for a given application. 

The framework builds upon previous MDP-based techniques by adding the concept 
of transition states, and demonstrates the utility of this concept in making 
effective system level reconfiguration decisions.  The framework also 
incorporates a novel grouping of relevant modeling constructs using the 
matrix Kronecker product, which can significantly reduce the computational 
time required for policy generation when developing large systems. 

We have developed our SLRF in general terms, and shown how to apply the general
framework to a specific application and embedded system implementation.  Since
the method is a formulaic approach, it can be used to capture and formalize
system knowledge that may otherwise exist only in the minds of engineers that
have spent years acquiring the knowledge through experience.  Such a
distillation can be useful, for example, when experienced engineers leave a
project, and take that intuition with them. 

Additionally, our SLRF provides an
automatic way to re-visit control decisions when a system property changes. For
example, if the access time of the off-chip non-volatile storage
in the EFM32GG system 
is improved due to new technologies or techniques, any manually
generated reconfiguration policies would need to be considered again to derive
new heuristics. In contrast, the MDP policies can be updated immediately, and
this updating can be done using direct formulas that generate optimal policies
based on the newly employed technologies or techniques.

We also note an alternative use for our proposed SLRF. If for any reason, a
system's designers are unable, unwilling or otherwise prevented from replacing
their manually generated controls with an MDP-generated
lookup table (LUT), they can use the
MDP policy as a comparative benchmark upon which to improve their existing
control algorithms. For example, the MDP and manual policies can be compared in
detail --- offline via inspection or through system simulations --- to identify
specific scenarios and conditions in which the heuristics and the MDPs conclude
different control decisions from the same observations. Those differences can
then be used to identify possible areas where the existing manual policies
might have room for improvement.

Other useful directions for future work include investigating the 
extension of our SLRF to incorporate multirate control policies, and building on the 
framework by extending from classical MDPs to partially-observable MDPs, which has 
the potential to further improve the utility of the framework.

\end {comment}

%% file: s07-ack.tex
This research was sponsored in part by the US National Science Foundation (CNS1514425 and CNS151304).